\documentclass[paper]{JHEP3} % 10pt is ignored!

\JHEPspecialurl{http://jhep.sissa.it/JOURNAL/JHEP3.tar.gz}

\usepackage{epsfig,multicol}

\newcommand\fverb{\setbox\pippobox=\hbox\bgroup\verb}
\newcommand\fverbdo{\egroup\medskip\noindent%
			\fbox{\unhbox\pippobox}\ }
\newcommand\fverbit{\egroup\item[\fbox{\unhbox\pippobox}]}
\newcommand {\beq}{\begin{equation}}
\newcommand {\eeq}{\end{equation}}
\newcommand {\beqa}{\begin{eqnarray}}
\newcommand {\eeqa}{\end{eqnarray}}
\newcommand {\n}{\nonumber \\}
\newcommand {\tr}{{\rm tr\,}}
\newcommand {\Tr}{\mbox{Tr\,}}

\newcommand {\Pf}{\mbox{Pf}}

\newcommand {\ee}{\mbox{e}}

\newcommand {\dd}{\mbox{d}}

\newcommand {\del}{\partial}

\newbox\pippobox
\title{Dynamical Generation of Four-Dimensional\\
Space-Time in the IIB Matrix Model}

\author{Jun Nishimura\\
The Niels Bohr Institute\\
Blegdamsvej 17, DK-2100 Copenhagen \O, Denmark, and\\
Department of Physics, Nagoya University\\
Furo-cho, Chikusa-ku, Nagoya 464-8602, Japan\\
E-mail: \email{nisimura@eken.phys.nagoya-u.ac.jp}}
\author{Fumihiko Sugino\\
Service de Physique Th\'{e}orique, C.E.A. Saclay\\
F-91191 Gif-sur-Yvette Cedex, France\\
E-mail: \email{sugino@spht.saclay.cea.fr}}
\preprint{\hepth{0111102}}	% OR: \preprint{Aaaa/Mm/Yy\\Aaa-aa/Nnnnnn}
\abstract{We study the spontaneous breakdown of SO(10) symmetry in the 
IIB matrix model, a conjectured nonperturbative definition
of type IIB superstring theory in ten dimensions. 
Our analysis is based on a Gaussian expansion technique, 
which was originally proposed by Kabat-Lifschytz and 
applied successfully to the strong coupling dynamics of the Matrix Theory.
We propose a prescription for including higher order corrections,
which yields a rapid convergence in a simple example.
This prescription is then applied to the IIB matrix model up to the
third order.
We find that the `self-consistency equations' allow 
various symmetry breaking solutions.
Among them, the solution preserving SO(4) symmetry
is found to have the smallest free energy.
The value of the free energy comes closer to the analytic formula
obtained by Krauth-Nicolai-Staudacher as we increase the order.
The extent of the space-time in the 4 directions is larger than 
the remaining 6 directions,
and the ratio increases with the order.
These results provide the first analytical evidence 
that four-dimensional space-time is generated dynamically 
in the IIB matrix model.
}


\begin{document} 

%\maketitle  IS IGNORED %%%%%%%%%%%

\section{Introduction}

One of the recent excitements in string theory is the
appearance of concrete proposals for its nonperturbative definitions.
The IIB matrix model \cite{IKKT},
which was conjectured to describe type IIB superstrings in 10 dimensions,
has a particularly simple form.
It is a supersymmetric matrix model, which can be obtained
from the zero-volume limit
%(a dimensional reduction to a point) 
of 10d SU($N$) super Yang-Mills theory.
%In principle, one should be able to predict the space-time dimensionality,
%the gauge group, and the number of generations, and so on.
The space-time is represented by 10 bosonic matrices,
and hence treated dynamically. 
This allows us in particular 
to investigate the possibility \cite{AIKKT}
that our 4d space-time is generated {\em dynamically}
in superstring theory in 10d.
Since the model has manifest SO(10) symmetry,
the emergence of 4d space-time requires the SO(10) symmetry
to be spontaneously broken.
%which requires the SO(10) symmetry to be spontaneous broken.
%Such a phenomenon should be associated with
%spontaneous breakdown of the SO(10) symmetry.
However, the absence of SSB has been concluded 
in various simplified versions
of the IIB matrix model
\cite{HNT,branched,4dSSB}. (See also Ref.\ \cite{Burda:2000mn}.)
These results suggested that the {\em phase} of the fermion integral  
must play a crucial role if the SSB really takes place.
A saddle-point analysis of its effect \cite{NV} led to 
the conclusion that the SO(10) symmetry can be spontaneously broken 
down to SO($d$), where $3 \le d \le 8$.
% The effects of the phase was studied using a saddle-point
% approximation \cite{NV}, it was concluded that the 
% SO(10) symmetry can be spontaneously broken down to SO($d$), 
% where $3 \le d \le 8$.
Monte Carlo studies have been attempted in Ref.~\cite{sign}
and an intuitive argument for $d=4$ has been presented.
Ref.~\cite{exact}, on the other hand, has provided
concrete examples of exactly solvable matrix models,
in which SO($D$) symmetry is spontaneously broken precisely
due to the phase.

In this paper, 
we address the issue of SSB in the IIB matrix model 
by using 
the Gaussian expansion technique 
developed in Refs.\ \cite{Oda:2001im,Sugino:2001fn}.
Such a method for general supersymmetric models
was originally proposed by Kabat-Lifschytz \cite{Kabat:2000hp} 
with the particular
intention of studying the strong-coupling regime 
of the Matrix Theory \cite{BFSS}.
Even at the {\em leading order} of the expansion,
the result turned out to be consistent with the conjectured
duality to supergravity, and various nonperturbative blackhole dynamics
have been extracted successfully \cite{blackholes}.
However, 
the possibility of a systematic improvement beyond the leading order
%the question of convergence 
was left unclear.
Here we will propose a prescription for including higher order 
corrections systematically, which indeed yields a rapid convergence 
to the exact result in a simple example.
% leads to a rapid convergence 
% which yields  
% and see in a simple example 
% that including higher order corrections leads to a rapid convergence 
% to the exact result. 
%discuss a simple example, in which rapid convergence
%is indeed observed.

We apply this prescription to the IIB matrix model
and carry out calculations up to the third order.
`Self-consistency equations' allow 
various solutions which preserve only some subgroup of SO(10).
% In the IIB matrix model, we find that `self-consistency equations' allow 
% various solutions which preserve only some subgroup of SO(10).
% We calculate the free energy for each solution up to the third order.
Among them, the solution preserving SO(4) symmetry
gives the smallest free energy.
% Among the solutions we have found, the free energy
% becomes minimum for the solution preserving SO(4) symmetry.
%For that solution, 
The value of the free energy comes much closer to the analytic formula
obtained by Krauth-Nicolai-Staudacher \cite{KNS}
as we increase the order.
As a fundamental observable which probes the space-time structure, 
we calculate the extent of space-time in each direction.
For the SO(4) preserving solution,
the extent in the four directions
%corresponding to the remaining SO(4) symmetry 
is found to be larger than the remaining six directions 
and moreover the ratio becomes larger as we go to higher order.
% We also calculate the extent of space-time in each direction.
% and observe that 
% the extent in the four directions
% %corresponding to the remaining SO(4) symmetry 
% is larger than the rest and moreover the ratio becomes larger
% as we go to higher order.
These results provide the first analytical evidence that
four-dimensional space-time is generated dynamically in the IIB matrix
model.
% support the possibility 
% that four-dimensional space-time is generated
% {\em dynamically} in the IIB matrix model.
%we provide a first analytical evidence that
%four-dimensional space-time is generated dynamically in the IIB matrix
%model.

%\setcounter{equation}{0}
\section{Bosonic Yang-Mills Integral}
%\label{bosonic}
In order to illustrate our method,
let us consider the bosonic Yang-Mills integral
defined by
%, which is given by the partition function
\beqa
Z &=& \int \dd A \, \ee ^{-S} \ ,
\label{bosonicZ} \\
S &=& - \frac{1}{4}N\beta \, \tr [A_\mu , A_\nu] ^2  \ ,
~~~~~~~~~~\beta=\frac{1}{g^2 N} \ .
\label{bosonicSA}
\eeqa
The bosonic matrices $A_\mu$ ($\mu = 1, \cdots , D$) 
are $N\times N$ hermitian matrices, which we expand as
$A_\mu = A_\mu ^a \, T^a $
with respect to the SU($N$) generators $T^a$ ($a=1,\cdots ,(N^2-1)$) 
normalized as $\tr (T^a T^b) = \frac{1}{2} \delta ^{ab}$.
The integration measure for $A_\mu$ is defined as
$ \dd A = \prod_{a=1}^{N^2-1} \prod_{\mu = 1}^{D}
\frac{d A_\mu ^a}{\sqrt{2 \pi}} $.
As an important consequence of the zero-volume limit,
one can actually absorb the parameter $g$
by rescaling the dynamical variables $A_\mu \mapsto \sqrt{g}A_\mu$.
Therefore, the parameter $g$ is merely a scale parameter rather than a
coupling constant.
The partition function is conjectured to be 
finite \cite{Krauth:1998yu}
for $N > D/(D-2)$, and this conjecture was proved 
in \cite{Austing:2001bd}.
A systematic $1/D$ expansion has been formulated in \cite{HNT}.
In particular the absence of SO($D$) breaking is shown 
to all orders of the $1/D$ expansion
and this conclusion is confirmed by 
Monte Carlo simulations \cite{HNT}
for various $D=3,4,6,\cdots ,20$.
The model has been studied by the Gaussian expansion
assuming that the SO($D$) symmetry is not spontaneously 
broken \cite{Oda:2001im},
and the numerical results of the VEVs of 
Polyakov line and 
Wilson loop \cite{AABHN}
have been reproduced qualitatively.

In order to examine the SSB of SO($D$) symmetry
by means of the Gaussian expansion,
we repeat the calculation of Ref.\ \cite{Oda:2001im}
without assuming the absence of SO($D$) breaking.
It turns out to be convenient to 
introduce the rescaled dynamical variables $X_\mu$ given by 
$
X_\mu = \beta ^{1/4} A_\mu \ ,
$
so that the action takes the canonical form
\beq
S = - \frac{1}{4}N \, \tr [X_\mu , X_\nu] ^2  \ .
\label{bosonicS}
\eeq
The most general SU($N$) invariant Gaussian action
without assuming SO($D$) symmetry 
can be brought into the form
\beq
S_0 = \sum _{\mu = 1} ^{D}  \frac{N}{v _\mu}
\tr \left( X_\mu X_\mu \right) \ ,  ~~~~~~~~~~v_\mu > 0  \ ,
\label{S0def}
\eeq
by making an appropriate SO($D$) transformation.
Then we rewrite the partition function (\ref{bosonicZ}) as
\beqa
Z &=& Z_0  \, \langle  \ee ^{- (S-S_0)} \rangle _0 \ , 
\label{factori}
\\
Z_0 &=& \int \dd A \, \ee ^{-S_0} = \beta ^{- D(N^2-1)/4}  
\int \dd X \, \ee ^{-S_0}  \ ,
\eeqa
where $\langle \ \cdot \ \rangle_0$ is a VEV with respect to the
partition function $Z_0$.
{}From (\ref{factori}) it follows that the free energy $F = - \ln Z$
can be expanded as
\beqa
F &=& \sum_{k=0}^{\infty} F_k  ~~~;~~~
F_0 = - \ln Z_0 \ , \n
F_k &=&  -  \frac{(-1)^k}{k!} \langle (S - S_0)^k \rangle_{{\rm C},0} 
~~~~~~~~(\mbox{for}~k\ge1) \ ,
\label{free_expand}
\eeqa
where the suffix `C' in $\langle \ \cdot \ \rangle _{{\rm C} , 0}$
means that the connected part is taken.
The first two terms of the expansion are given as
\beqa
F_0 &=&  \frac{1}{2} (N^2 -1 ) 
\Bigl\{ D \ln (N \beta^{1/2})
-  \sum _{\mu = 1} ^D \ln v _\mu  \Bigr\} \ , \\
F_1 &=& \langle   S  \rangle_0
- \langle   S_0 \rangle_0  \ , \\
\label{SB0}
\langle   S  \rangle_0
&=& \frac{1}{8} (N^2 -1)
\sum _{\mu \ne \nu} v _\mu v_\nu \ ,  \\
\langle   S_0 \rangle_0
&=& \frac{1}{2} (N^2 -1) D \ .
\eeqa

%The usual mean-field approximation amounts to choosing 
The variational parameters $v _\mu$ 
in the Gaussian action (\ref{S0def})
can be determined in such a way that the free energy $F$ calculated 
up to the first order becomes minimum.
This gives the self-consistency equations 
\beq
\label{gapeq}
0 = \frac{1}{N^2 -1} \frac{\del}{\del v _\mu} (F_0 + F_1) 
  = - \frac{1}{2 v_\mu} +  \frac{1}{4} \sum _{\nu \ne \mu} v_\nu  \ .
\eeq
Considering $v_\mu  > 0$, one immediately finds that
\beq
v_1 = v_2 = \cdots = v_D = 
\sqrt{\frac{2}{D-1}} \ ,
\label{v_bosonic}
\eeq
which agrees with Ref.\ \cite{Oda:2001im}.
Thus the Gaussian approximation is able to reproduce correctly
the absence of the SO($D$) symmetry breaking.

\section{The IIB matrix model}

\label{IIB}
Let us move on to the IIB matrix model, which is defined by the 
partition function
\beq
Z = \int \dd A \dd \Psi
\, \ee ^{- S }  ~~~;~~~
S = S^{(\rm B)} + S^{(\rm F)}  \ ,
\label{susyZ}
\eeq
where $S^{(\rm B)}$ is the bosonic action given by (\ref{bosonicSA}) 
with $D=10$,
and $S^{(\rm F)}$ is given by
\beq
S^{(\rm F)}
 = - \frac{i}{2}N \beta \,  \tr \Bigl( \Psi_\alpha
(\widetilde{\Gamma}_\mu)_{\alpha\beta} [A_\mu , \Psi _\beta ]  \Bigr) \ .
\label{fermionic_action}
\eeq
The fermionic matrices 
$\Psi _\alpha$ ($\alpha = 1 , \cdots , 16$)
are traceless $N \times N$ hermitian matrices,
%and we expand them as 
which we expand as 
$\Psi _\alpha = \Psi_\alpha ^a \, T^a $.
The integration measure is given by
$\dd \Psi 
= \prod _{a=1}^{N^2-1} \left(
\prod _{\alpha=1}^{16}  \dd \Psi_\alpha ^a \right) $.
The $16\times 16$ symmetric matrices $\widetilde{\Gamma}_\mu$ are 
given as
$\widetilde{\Gamma}_\mu = {\cal C}\, \Gamma_\mu$,
where $\Gamma_\mu$ are the 10d gamma matrices after Weyl projection,
and ${\cal C}$ is the charge conjugation matrix, which satisfies
$\Gamma_\mu^\top =  {\cal C} \Gamma_\mu  {\cal C}^\dag$
and ${\cal C}^\top = {\cal C}$.
The partition function (\ref{susyZ})
is conjectured \cite{KNS} to be finite for arbitrary $N$, 
%(and also for obvious generalizations to $D=4$ and 6), 
and this conjecture was proved in \cite{AW}.

Again we introduce the rescaled variables $X_\mu$,  
%by (\ref{Xdef}),
and similarly we introduce $\Phi_\alpha$ by
$
\Phi_\alpha = \beta^{3/8} \,  \Psi_\alpha \ , 
$
so that the fermionic action takes the canonical form 
\beq
S^{(\rm F)}
 = - \frac{i}{2}N  \,  \tr \Bigl( \Phi_\alpha
(\widetilde{\Gamma}_\mu)_{\alpha\beta} [X_\mu , \Phi _\beta ]  \Bigr) \ .
\label{fermionic_action_can}
\eeq
We write down the SU($N$) symmetric
Gaussian action without assuming SO(10) invariance as
\beq
S_0 = S_0 ^{(\rm B)} + S_0 ^{(\rm F)}  \ , 
\label{totalaction}
\eeq
where $S_0 ^{(\rm B)}$ is the anisotropic 
bosonic Gaussian action (\ref{S0def})
and $S_0 ^{(\rm F)}$ is written as
\beq
S_0 ^{\rm (F)} = 
\frac{N}{2}
\sum_{a=1}^{N^2-1} \Phi _\alpha ^a \, 
{\cal A} _{\alpha\beta} \, \Phi _\beta ^a  \ ,
\label{SFgauss}
\eeq
where ${\cal A}$ is a $16\times 16$ anti-symmetric matrix.

As before, we introduce $Z_0$, $\langle \ \cdot \ \rangle_0$ 
and the free energy is expanded as (\ref{free_expand}).
Note that correlation functions 
$\langle \ \cdot \ \rangle_{{\rm C},0}$
including odd powers of $S^{\rm (F)}$ vanish trivially.
When we evaluate the free energy $F$ by using 
the formula (\ref{free_expand}),
we have to reorganize the expansion in such a way that
we can take into account the three-point interaction $S^{\rm (F)}$ 
properly.
As in Ref.\ \cite{Sugino:2001fn}, we consider
$S^{\rm (B)}$,
$S_0^{\rm (B)}$, $S_0^{\rm (F)}$, $(S^{\rm (F)})^2$ to be of the same
`order' and arrive at the new expansion
\beqa
F &=& \sum_{k=0}^{\infty} \tilde{F}_k \ ;~~~~~~
\tilde{F}_0 = - \ln Z_0  \ ;
\label{free_expand2def}
\\
\tilde{F}_k &=&  
- \sum_{l=0}^{k} a_{k,l} \, 
\Bigl\langle (S^{\rm (B)} - S_0 )^{k - l} (S^{\rm (F)})^{2l}
\Bigr \rangle_{{\rm C},0} \ ,  \n
&~& a_{k,l}=  \frac{(-1)^{k+l}}{(k+l)!} ~_{k+l} \mbox{C} _{k-l} 
~~~~~~~~(\mbox{for}~k\ge1) \ .
\label{free_expand2}
\eeqa
This prescription was first proposed in Ref.\ \cite{Kabat:2000hp}
for general supersymmetric models 
\footnote{
\label{auxiliaryfields}
In Ref.\ \cite{Kabat:2000hp} the Gaussian action is 
constructed using a superfield formalism.
Similarly we may use the four-dimensional superfield formalism
regarding the IIB matrix model as the zero-volume limit 
of 4d ${\cal N}=4$ super Yang-Mills theory.
%This prescription preserves SO(4) subgroup of SO(10).
The result turns out to be 
%qualitatively the same 
quite similar 
to what we obtain below
for the solution assuming the SO(4) symmetry.
%In fact, we checked that for the SO(4) case  
%the result here does not show essential differences 
%from the result of the superfield formalism up to the order 1.
}.

The zeroth order free energy can be obtained as
\beqa
\label{Fzeroth}
\tilde{F}_0 &=& (N^2 -1 ) \left\{
C - \frac{1}{2} \sum _{\mu = 1} ^{10} \ln v _\mu 
-  \ln  (\Pf \, {\cal A})   
\right\} \ ,  \\
\label{defC}
C &\equiv& \frac{1}{2} \Bigl\{  10  \ln (N \beta ^{1/2})
- 16 \ln (N \beta ^{3/4}) \Bigr\} \ .
\eeqa
The first order correction to the free energy reads
\beq
\tilde{F}_1 = \langle   S^{\rm (B)}  \rangle_0
- \langle   S_0 \rangle_0
- \frac{1}{2} \Bigl\langle (S^{\rm (F)})^2 \Bigr\rangle _{{\rm C} , 0} \ .
\eeq
The first term is the same as (\ref{SB0}) in the bosonic case.
The other terms are given as
\beqa
 & & \langle   S_0 \rangle_0 =  - 3  (N^2 -1) \ , \\
 & & \langle (S^{\rm (F)})^2 \rangle _{{\rm C} , 0}
= \frac{1}{2 } (N^2-1)  Q \ ,
\label{Sf2} \\
 & & Q \equiv  \sum_\mu \rho_\mu v_\mu \ ,
~~~~~
\rho _\mu \equiv  \frac{1}{4} \Tr
 \Bigl\{ ( {\cal A}^{-1} \tilde{\Gamma}_\mu )^2 \Bigr\}   \ ,
\label{defrho} 
\eeqa
where the trace $\Tr$ is taken with respect to the $16$-dimensional
spinor indices.
Thus, at the first order, the free energy is calculated as
\beqa
\frac{1}{N^2-1} (\tilde{F}_0 + \tilde{F}_1)  
& = & C  -   
\frac{1}{2}\sum _{\mu = 1} ^{10} \ln v _\mu  
- \ln (\Pf \, {\cal A})  \n 
 & &  + \frac{1}{4} \sum _{\mu \ne \nu} v _\mu v_\nu  
+ 3 - \frac{1}{4}  Q  \ .
\label{calFexplicit}
\eeqa

Let us parametrize the $16 \times 16$ anti-symmetric matrix
${\cal A} _{\alpha\beta}$ in a SO(10) covariant way as
\beq
{\cal A} = \frac{i}{ 3 !} 
\sum_{\mu \nu \lambda}
w_{\mu\nu\lambda} \, {\cal B}_{\mu\nu\lambda}
\ ; ~~~~~
{\cal B}_{\mu\nu\lambda}
={\cal C} \,
\Gamma_\mu  \, \Gamma_\nu^\dag \,  \Gamma_\lambda \ ,
\label{w_def}
\eeq
where $w_{\mu\nu\lambda}$ is a complex totally anti-symmetric
rank-three tensor.
As it becomes clear from this parametrization, 
the fermionic Gaussian action breaks
the full SO(10) symmetry for any nonzero $w_{\mu\nu\lambda}$.
The self-consistency equations at the first order read
\beqa
0 &=& -\frac{1}{2 v_\mu} 
+ \frac{1}{4} \sum _{\nu \ne \mu} v_\nu
- \frac{1}{4} \rho_\mu \ , 
\label{self_v}
\\
0 &=& - \frac{1}{2} \, \Tr ({\cal A}^{-1} {\cal B}_{\mu\nu\lambda} )
%\n 
%&~& 
+ \frac{1}{8} \sum_\mu v_\mu 
\Tr \Bigl\{ ( {\cal A}^{-1}  \widetilde{\Gamma}_\mu )^2
{\cal A}^{-1} {\cal B}_{\mu\nu\lambda}   \Bigr\} \ . 
\label{self_A}
\eeqa

%\setcounter{equation}{0}
%\section{Systematic higher order calculations}
%\label{higher}

In order to go beyond the leading order approximation,
we specify the prescription for 
systematic higher order calculations as follows.
We choose the parameters in the Gaussian action
in such a way that
the truncated free energy $\sum_{k=0}^n \tilde{F}_k$
is extremized.
%up to the $n$-th order 
%This gives the self-consistency equations at the $n$-th order.
(Technically, we search for solutions by Newton's method 
near the solutions found at lower orders.)
This will give the $n$-th order approximation 
to the free energy.
Similarly to the free energy (\ref{free_expand2def}),
the expectation value of
an operator ${\cal O}$ is calculated using the
reorganized series expansion
\beqa
\label{operator_inf}
\langle {\cal O} \rangle &=& \langle {\cal O} \rangle_0 + 
\sum_{k=1}^{\infty}
\tilde{O}_k
%\langle {\cal O} \rangle^{(n)} 
\ ; \\
%\langle {\cal O} \rangle^{(n)} 
\tilde{O}_k
& \equiv &  
\sum_{l=0}^{k} a_{k,l} \, 
\Bigl\langle (S^{\rm (B)} - S_0 )^{k - l} (S^{\rm (F)})^{2l}
{\cal O}
\Bigr \rangle_{{\rm C},0} \ .
\label{operator_k}
\eeqa
We truncate the infinite series (\ref{operator_inf})
at $k=n$,
%at the $n$-th order,
and evaluate it at the solutions to the
$n$-th order self-consistency equations.
%\cite{higherorder_gapeq}.  
%\bibitem{higherorder_gapeq}

As a test, we applied this prescription
to a toy model \cite{Oda:2001im} with the action
$S=\frac{1}{4g^2}\phi^4$.
Here we can easily proceed up to the 10th order (or even higher)
by using Mathematica.
We found solutions to the self-consistency equation
except for the 2nd and 4th orders.
At each order (other than 2 and 4),
we calculated the free energy and the expectation value
$\langle \phi^2 \rangle$.
For the free energy, 
the discrepancy from the exact result is 
only 0.1\% already at order 3, and 
it becomes 0.004\% at order 10.
For the expectation value $\langle \phi^2 \rangle$,
the discrepancy is found to be 0.4\% already at order 3, and 
it becomes 0.03\% at order 10.
This suggests that the above prescription for higher order calculations
indeed yields a rapid convergence.\footnote{
We have also tried other prescriptions.
For instance, when we evaluate the expansions
%(\ref{free_expand2def}) and (\ref{operator_inf})
truncated at the $n$-th order,
we may use the solution to the {\em first}
order self-consistency equations.
However, the results obtained for the free energy and 
$\langle \phi^2 \rangle$ both start oscillating at the third order,
and the oscillation becomes increasingly violent as we go to higher order.} 
Details shall be reported elsewhere.

\section{Ansatz}
\label{ansatz}
Since the Gaussian action contains too many parameters 
(10 real numbers from $v_\mu$ in (\ref{S0def}) and
120 complex numbers from $w_{\mu\nu\lambda}$ in (\ref{w_def})),
it seems a formidable task to explore
the whole solution space of 
the self-consistency equations.
% (\ref{self_v}), (\ref{self_A}).
Here we search for solutions
assuming that SO($d$) plus some discrete subgroup 
of SO(10) is preserved.
For each case below ($d = 2, 4, 6, 7$)
the parameters are reduced to two real and one complex numbers. 
For other values of $d$, we need to keep more independent parameters,
and we leave these cases for future investigations.

First let us assume that SO(7) symmetry is preserved,
which allows us to set $v_1 = \cdots = v_7 \equiv V$
and all the $w_{\mu\nu\lambda}$ except $w_{8,9,10} \equiv w$ to zero.
By further imposing the symmetry under cyclic permutations
of $x_8$, $x_9$ and $x_{10}$, 
we set $v_8 = v_9 = v_{10}\equiv v$.

Secondly let us assume SO(6) symmetry,
which allows us to set $v_1 = \cdots = v_6 \equiv V$
and all the $w_{\mu\nu\lambda}$ with indices 
$1 \sim 6$ to zero.
%We end up with four parameters $w_{8,9,10}$, $w_{7,9,10}$, $w_{7,8,10}$ and
%$w_{7,8,9}$ .
We further impose the symmetry under even permutations
of $x_7$, $x_8$, $x_9$ and $x_{10}$. 
This requires $v_7 = v_8 = v_9 = v_{10} \equiv v$
and 
$w_{7,8,9} = - w_{7,8,10} = w_{7,9,10} = - w_{8,9,10} \equiv w$.

Next let us assume SO(4) symmetry, which allows
us to set $v_1 = \cdots = v_4 \equiv V$
and all the $w_{\mu\nu\lambda}$ with indices 
$1 \sim 4$ to zero.
We also impose the symmetry 
under a particular SO(10) transformation
\beq
x_\mu \mapsto x_\nu  ~~~;~~~ x_\nu \mapsto x_\mu ~~~;~~~x_1 \mapsto - x_1 \ ,
\label{discrete2SO10}
\eeq
where $(\mu , \nu)=(5,6),(7,8),(9,10)$. 
Furthermore we impose the symmetry under cyclic permutations
of the three pairs 
$(x_5, x_6), (x_7, x_8), (x_9, x_{10})$. 
%$(5,6),(7,8),(9,10)$.
This leads to $v_5 = \cdots  = v_{10} \equiv v$ and 
$w_{\mu\nu\lambda} \equiv w$, 
where $\mu \in \{5,6\}$, $\nu \in \{7,8\}$, 
$\lambda \in \{9,10\}$.
% $w_{5,7,9} = w_{5,7,10} = w_{5,8,9} = w_{5,8,10}=
%  w_{6,7,9} = w_{6,7,10} = w_{6,8,9} = w_{6,8,10} \equiv w$. 
% This requires $v_5 = \cdots  = v_{10} \equiv v$
% and $w_{8,9,10} = w_{5,9,10} = w_{8,6,10} = w_{8,9,7}=
% w_{5,6,10} = w_{5,9,7} = w_{8,6,7} = w_{5,6,7}$.

Finally let us assume SO(2) symmetry, which allows us to set
%Then we can set
$v_1 = v_2  \equiv V$
and all the $w_{\mu\nu\lambda}$ with indices 
$1$ or $2$ to zero.
We also impose the symmetry 
under the transformation (\ref{discrete2SO10})
with $(\mu , \nu)=(3,4),(5,6),(7,8),(9,10)$.
%with $(\mu , \nu)=(3,7),(4,8),(5,9),(6,10)$. 
Furthermore we impose the symmetry 
under even permutations of the four pairs 
$(x_3, x_4), (x_5, x_6), (x_7, x_8), (x_9, x_{10})$. 
%the transformation
%(\ref{discreteSO10})
%with $(\mu , \nu , \lambda)=(7,8,9),(7,8,10),(7,9,10),(8,9,10)$
This requires $v_3 = \cdots  = v_{10} \equiv v$ and 
$w_{\mu \nu \lambda} = -w_{\mu\nu\rho} = w_{\mu \lambda \rho} = 
-w_{\nu \lambda \rho} \equiv w$, where $\mu \in \{3,4\}$, $\nu \in \{5,6\}$, 
$\lambda \in \{7,8\}$, $\rho \in \{9,10\}$. 
% This requires $v_3 = \cdots  = v_{10} \equiv v$
% and 
% $w_{8,9,10} = - w_{7,9,10} = w_{7,8,10} = - w_{7,8,9}
% = w_{4,9,10} = w_{8,5,10} = w_{8,9,6} = w_{4,5,10} = w_{4,9,6} =
% w_{8,5,6} = w_{4,5,6} $

%\setcounter{equation}{0}
\section{Results}
\label{results}
For each ansatz
preserving the SO($d$) symmetry ($d=2,4,6,7$), 
%we obtain a solution or two.
we obtain one or two solutions.
In the latter case, we only show results for the one which gives
the smaller free energy.
An analytic formula for the free energy at arbitrary $N$
was obtained in Ref.~\cite{KNS} 
% based on previous analytical calculations \cite{Moore:2000et} 
% and explicit numerical evaluations at small $N$.
based on explicit numerical evaluations at small $N$
combined with other analytical calculations \cite{Moore:2000et}.
At large $N$, the formula gives \footnote{The 
famous factor \cite{Moore:2000et}
$\sum_{n|N} \frac{1}{n^2}$ in the partition function 
gives an $O(N^{-2})$ contribution in eq. (\ref{exact_F}),
hence it is irrelevant in the present analysis.} 
\beq
\frac{F}{N^2 -1}  = 
 \ln (\sqrt{N}g^7) + \left(\ln 8 - \frac{3}{4}\right) 
 + O\left(\frac{\ln N}{N^2} \right) \ . 
\label{exact_F} 
\eeq
% \beqa
% \frac{1}{N^2 -1}F & =  &
%  \ln (\sqrt{N}g^7) + \left(\ln 8 - \frac{3}{4}\right)  \n 
% &~& + \frac{11}{12N^2}\ln N + O\left(\frac{1}{N^2}\right) \ .
% \eeqa
The Gaussian expansion reproduces the first term
correctly for any solution.
Therefore, we compute the `free energy density' defined by
\beq
f = \lim _{N\rightarrow \infty} \Bigl\{
\frac{1}{N^2 -1}F - \ln (\sqrt{N}g^7) \Bigr\} \ .
\eeq
The results are shown in the Table. 
\TABLE{
\begin{tabular}{|c|c|c|c|c|}
\hline\hline
$d$ & $f$ (order 1) & $f$ (order 3) & $\rho$ (order 1) & $\rho$ (order 3)\\
\hline
2 & 6.49428  & 6.50906 & 2.17736 & 1.49056 \\
4 & 6.15335  & 0.74111 & 1.85728 & 3.37766 \\
6 & 5.75743  & 1.54414 & 1.87034 & 2.24911 \\
7 & 5.52272  & 1.62094 & 1.95533 & 2.15681 \\
\hline\hline
\end{tabular}
}
At the first order, the free energy becomes larger for smaller $d$. 
At the second order, we find no solutions to the self-consistency
equations. This is not so surprising, however, 
since we encounter a similar situation with the $\phi^4$ toy model 
as we mentioned below (\ref{operator_k}).
%It is expected that solutions persist to exist for sufficiently high
%orders.
At the third order,
we find that the free energy becomes minimum at the solution
preserving SO(4) symmetry.
%Note also that the minimum value 
%is close to the `exact' result ($\ln 8 - \frac{3}{4}= 1.32944$).
Note also that the value of $f$ obtained for $d=4$ 
comes much closer
to the `exact' result ($\ln 8 - \frac{3}{4}= 1.32944$)
as we proceed from order 1 to order 3.
% at the third order is much closer 
% to the `exact' result ($\ln 8 - \frac{3}{4}= 1.32944$)
% than the result obtained for $d=4$ at the first order.
%
% Note also that the 
% minimum value obtained at the third order (0.74111)
% is much closer to the `exact' result ($\ln 8 - \frac{3}{4}= 1.32944$)
% than the minimum value obtained at the first order (5.52272).
We also calculate the extent in the $\mu$-th direction
$R_{\mu} \equiv \sqrt{\langle \frac{1}{N}\tr (A_{\mu})^2 \rangle}$.
%using the approximation.
Note that $R_1 = \cdots = R_d \equiv R$
and $R_{d+1} = \cdots = R_{10} \equiv r$
due to the imposed symmetry.
At the first order, the ratio $\rho \equiv R/r$ is given by 
$\sqrt{V/v}$ and we find that $\rho > 1$.
At the third order, we observe that the ratio $\rho$ increases 
in all the cases except for $d=2$.

\section{Discussion}
\label{summary}

In this paper we have formulated an analytical approach to
the spontaneous breakdown of SO(10) symmetry
in the IIB matrix model.
Our approach is based on the Gaussian expansion technique,
which was quite successful in the Matrix Theory
even at the leading order.
We have given a prescription for systematic higher order
calculations, which, to our surprise, 
%converge rapidly 
yields a rapid convergence 
in a simple example.
%We are currently studying the convergence in a more nontrivial model.

Here we have carried out our program up to the third order,
which can be done with reasonable efforts 
({\em e.g.}, we evaluated 34 four-loop diagrams).
We found various solutions which break SO(10) symmetry,
and among them the SO(4) preserving solution turned out to
give the smallest free energy.
Moreover the ratio 
%$\rho = R/r$ 
of the extents in the
4 and the other 6 directions increases as we go to higher order.
These results support 
the conjectured scenario that 4d space-time is generated dynamically
in nonperturbative superstring theory.
%which is in accord with an intuitive argument
%based on low-energy effective theory and recent Monte Carlo results .
Let us also recall that 
the emergence of four-dimensional space-time in the IIB matrix model
has already been suggested in Ref.~\cite{sign} based on Monte Carlo
results and the branched polymer
description \cite{AIKKT} of its low-energy effective theory.
It is encouraging that we arrive at `4d' 
%is obtained in the present work 
from a totally different approach.

The analytic
formula for the free energy
obtained by Krauth-Nicolai-Staudacher \cite{KNS} 
provides a useful guide-line for the convergence of the present
approach.
Indeed we observed that the free energy calculated for the 
SO(4) preserving solution comes much closer to the KNS result
as the order is increased.
In the bosonic model, the Gaussian expansion (even at the first
order) becomes {\em exact} in the large-$D$ limit \cite{Oda:2001im}.
This fact can be understood naturally from the viewpoint of 
a systematic $1/D$ expansion \cite{HNT}.
In the supersymmetric models, $D$ is restricted to 4, 6 and 10
and therefore `the large-$D$ limit' does not make sense.
%it is not possible to formulate a $1/D$ expansion.
Still it is conceivable that the convergence is reasonably fast
if $D$ is as large as 10, {\em i.e.} 
the case corresponding to the IIB matrix model.
%the Gaussian expansion is expected to converge faster
%for larger $D$.
%It would be interesting to proceed to higher orders to confirm the 
%convergence.
The same reasoning explains the success of the (leading-order) 
Gaussian approximation in the Matrix Theory \cite{blackholes}.
Nevertheless it is certainly 
worth while to perform higher order calculations
and confirm the convergence directly.
Incidentally we had to restrict the parameter space of the Gaussian action
for a purely technical reason.
It would be interesting to enlarge the parameter space to study
other SO($d$) preserving solutions, in particular $d=3$ and $d=5$.

%At order 3, we had to evaluate 37 four-loop diagrams

%Also, it is interesting to extend our analysis to higher orders 
%and see whether the results are stable as in the $\phi^4$ toy model. 

Finally we would like to comment on the scale parameter $g$
in the action (\ref{bosonicSA}) and (\ref{fermionic_action}).
In the present approach, we rescale the matrices
in such a way that the action takes the canonical form
(\ref{bosonicS}), (\ref{fermionic_action_can}).
This corresponds to setting $g^2 N = 1$ in the original action.
Then the $N$ dependence is eliminated from the self-consistency
equations.
If $\langle {\cal O} \rangle$ converges 
%to a finite quantity
in high order calculations,
it means that the quantity becomes finite in the large-$N$ limit 
with $g^2 N$ fixed.
This explains the observed large-$N$ behavior of the
space-time extent and the Wilson loops
in the bosonic and supersymmetric models \cite{HNT,AABHN}.
In the IIB matrix model, it is therefore plausible that
$g^2 N$ should be fixed in order to obtain finite Wilson loop
correlators in the large-$N$ limit.
On the other hand, Monte Carlo results \cite{sign} suggest that
the extent of the space-time (in the 4 directions)
may diverge in this large-$N$ limit.
It would be interesting to see if such trends appear
in the higher order calculations.
% Monte Carlo results suggest that the ratio can be very large or
% We have found that the ratio $\rho = R/r$ of the extents in the
% 4 and 6 directions increases as we go to higher order.
% Monte Carlo results suggest that the ratio can be very large or
% even diverge \cite{sign}.
% %It is possible that the extent $R$ in the four directions actually
% %diverge
% Therefore we consider it possible that the ratio $\rho$ may become
% much larger as we go to higher order.
We hope that our analytical approach to the IIB matrix model
is useful to extract more information on its highly nontrivial dynamics.
%the supersymmetric Yang-Mills integrals
%such as 
%the IIB matrix model and the Matrix Theory.

\bigskip

\acknowledgments

The authors would like to thank I.\ Kostov, P.\ Di Francesco
and other organizers of the Sixth Claude Itzykson Meeting
(Matrix Models 2001), which triggered the current 
collaboration.

\end{document}